\documentclass[twocolumn,showpacs,preprintnumbers,prd,nofootinbib]{revtex4-1}
\usepackage{latexsym}
\usepackage{graphicx}

\begin{document}
\title{Spatial and temporal dynamics of infected populations: the Mexican epidemic}
\author{Mario A. Rodr\'iguez-Meza}
\affiliation{Departamento de F\'{\i}sica, Instituto Nacional de
Investigaciones Nucleares, Apdo. Postal 18-1027, M\'{e}xico D.F. 11801,
M\'{e}xico \\ 
e-mail: marioalberto.rodriguez@inin.gob.mx%; 
%web page: http://www.astro.inin.mx/mar
}

\begin{abstract}
Recientemente apareci\'o en M\'exico y en otras naciones la pandemia
asociada al virus A/H1N1-2009.
Presentamos un estudio de esta pandemia en el caso de M\'exico usando
el modelo SIR de epidemias. Este modelo es uno de los modelos m\'as
simples pero ha tenido \'exito en epidemias para poblaciones cerradas.
Consideramos los datos para M\'exico y usamos el modelo SIR para
hacer algunas predicciones. Despu\'es consideramos la generalizaci\'on
del modelo SIR para obtener adicionalmente el comportamiento 
espacial de la epidemia. Hacemos el estudio de la propagaci\'on
espacial y temporal de la epidemia con par\'ametros que son consistentes
con los obtenidos de ajustes del modelo SIR temporal al caso
de M\'exico.

\medskip
\noindent
\textit{Descriptores}: 
Din\'{a}mica de epidemias; modelo SIR espacial; ecuaciones de raz\'{o}n 
no-lineales.

\bigskip
Recently the A/H1N1-2009 virus pandemic appeared in Mexico and in other
nations. We present a study of this pandemic in the Mexican case using
the SIR model to describe epidemics. This model is one of the simplest
models but it has been a successful description of some epidemics of
closed populations. We consider the data for the Mexican case and use
the SIR model to make some predictions. Then, we generalize the SIR
model in order to describe the spatial dynamics of the disease. We make
a study of the spatial and temporal spread of the infected population
with model parameters that are consistent with temporal SIR model 
parameters obtained by fitting to the Mexican case.

\end{abstract}

\date{\today}
\preprint{}
\maketitle

%%%%%%%%%%%%%%%%%% INTRO %%%%%%%%%%%%%%%%%%

\section{Introduction}
The human being has suffered several pandemics along its history. 
In the last
century one of the most devastatings was the spanish flu that 
killed 
more 
people
than the world war I itself. 
It is important to know its dynamics to take appropriate
health measures in order to diminish the effects of the pandemics. 
By using population dynamics methods we can model 
the spatial and temporal dynamics of the disease\cite{Murray1989}.

The purpose of this work is to present a simple model to study the 
epidemic dynamics based on rate equations of the populations dynamics.
The model was developed by Kermack and McKendrick\cite{Kermack1927}
and it is known as the Susceptible-Infected-Released (SIR) model.
We have used the SIR model and one of its possible extensions
to take into account the oscillatory behavior of the infected population
 to study the A/H1N1-2009 pandemic in the case of Mexico\cite{Mar2010}. 
One of the disadvantages of 
the SIR model is that it does not take into account
for the fact 
that the infected people 
have the greatest influence on the healthy people
that are spatially closest to them. This also means that the SIR model ignores
the possibility that populations are spread over a spatial region and that it takes
time for an infection to spread across that region.
We extend the SIR model in order to study the spatial behavior.

In the following section, Sec. \ref{SIR_model_sec}, we review very briefly the SIR model and apply it to
the case of the Mexican epidemic in Sec. \ref{Mexican_epidemic_sec}. 
Then in section \ref{Spatial_sec} we improve the SIR model in order to take into account
for the spatial behavior and apply it to the Mexican epidemic, using the SIR parameters values
found previously  in the simple SIR model. Our conclusions are in Sec. \ref{conclusions}.

%%%%%%%%%%%%%%%%% SECTION %%%%%%%%%%%%%%%%%%
\section{The SIR model}\label{SIR_model_sec}

This model considers that the population is divided
into
susceptible people ($S$), 
healthy
people that can be infected, the infected people ($I$),
and the released people ($R$), persons that were ill and then died or cured and then
become immune.
 The released people, 
also named, removed people can be those infected people
that are isolated.
We follow Ref. \cite{Mar2010} and refer the reader to this reference for more
details.

The basis of the model are: (a) There is only one infection responsible to
produce an infection in a host. (b) The end of the disease is the death or
a complete immunity. (c) The rate at which the disease propagates is
proportional to the product of susceptible and infected people, the proportionality 
factor is $a$, its value will give us the infection period (see below).
(d) All individuals are equally susceptible. (e) The system is closed, i.e.,
there are no births or migrations. (f) The total population $N$ is large
enough in order to have a deterministic description, i.e., based 
on its history. (g) The incubation period is short, i.e., a susceptible that
become ill can pass the virus to another susceptible. (h) The decreasing
speed of the infected people is proportional to its size, with the proportionality factor
$b$, the inverse of the released period. The model also assumes
that all the populations are homogeneously  mixed, they have equal
probability to be in contact between each other, except the isolated
or dead people.

Taking into account all these considerations and using normalized
variables to the total population $N$,
$s\equiv S/N$, 
$i\equiv I/N$, $r\equiv R/N$, 
the model is described by the non-linear differential equations:
\begin{eqnarray}
\frac{d s}{d t} &=& -s(t) i(t) \, , 
\label{SIR_eq1} \\
\frac{d i}{d t} &=& s(t) i(t) -\rho i(t) \, ,
\label{SIR_eq2} \\
\frac{d r}{d t} &=& \rho i(t) \, ,
\label{SIR_eq3}
\end{eqnarray}
where the constant
$\rho\equiv b/a N$. Time is in units of $1/Na$, i.e., gives us the time
in units of the infection period.

The first equation above indicates the rate at which the susceptible population
diminishes due to that they are being infected, according to assumption (c). The second
equation gives us the speed at which the infected growths according to
assumptions (c) and (h). The negative term in this equation is the one we
have to subtract according to assumption (h), when some members of the 
compartment $i$ are being move to compartment $r$.
Finally, the increasing rate of the removed people is contemplated in the
third equation according to assumption (h). 

The mathematical formulation of the epidemiological model is complete when
we fix the initial condition of the populations, for example, at time $t=0$,
that mark the start of the epidemic:
$s(t=0) = s_0 >0 ,
i(t=0) = i_0 >0,
r(t=0) = r_0 =0$.
We also add the condition that the system is closed, 
$s(t)+i(t)+r(t)=1$, i.e., the total population is constant. 
This condition, the initial values of the populations, and Eqs. (\ref{SIR_eq1}-\ref{SIR_eq3}) mean
that we need only the values
of two parameters, $s_0$ and $\rho$.

We should notice that this is a very simple model. The SIR model has been
modified or improved after it was proposed, in order to be applied to
specific situations \cite{Brauer,Ng,Liu,Shi}.
The models mainly have relaxed the (e) and (g) assumptions, given above,
introducing more complicated dynamics. For example, latency period of the 
VIH virus is very variable. There are ill people that show their first symptom a few
years later after they were infected and there exist cases where there is no
manifestation ever.
Besides, in an epidemic like AIDS there are births of infected and non-infected
children and so  the total number of the population is changing.

Other variations of the standard SIR model consider the case of epidemics
where the populations is heterogeneous with very susceptible groups as can be
the very young children or the very old people (see for instance \cite{Mollison}).

Even though the standard SIR model is simple, it allows us to make some
predictions about the initial behavior of the spread of an epidemic. When
a population is isolated or it is very weakly interacting with its surroundings,
the SIR model gives appropriate results\cite{Mar2010}.

%%%%%%%%%%%%%%%%% SECTION %%%%%%%%%%%%%%%%%%
\section{Pandemic by the A/H1N1-2009 virus}\label{Mexican_epidemic_sec}

In the spring of 2009 it appeared a new epidemic with unknown origin. One of the 
first cases was reported on april 12th, a woman 39 years old. She died the 
next day. By the end of april the Mexican government announced that 
emergency  measures have to be taken and schools were closed.

\begin{figure}[htbp]
\includegraphics[width=3.2in]{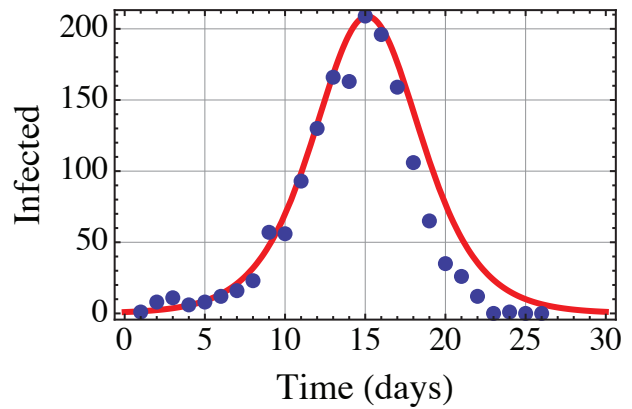}
\caption{Fitting of data reported on  the 8th of may.}
\label{fig01_May_8_plot}
\end{figure}

We applied in this section the SIR model 
to study
the pandemic by A/H1N1 using the available data for Mexico.
We 
show the results we obtained using the simplest SIR model\cite{Mar2010}.
The data we used were obtained from the web page of the 
Secretar\'\i a de Salud de Mexico\cite{Salud}.
In Fig. \ref{fig01_May_8_plot} we show the fitting results using the data
published on may 8th. The fitted parameters are: $Na=62$, $\rho=0.9928$,
$s_0=0.99999988$. The onset of the epidemic is determined by parameter
$\rho$ (see \cite{Mar2010} for details). This parameter is the most important in the
SIR model and not only determines the onset of the epidemic, it
determines the maximum of the infected people, i.e., how drastic is the epidemic.
As we defined this parameter $\rho=b/aN$, i.e.,  is the ratio of infection period over 
 released period.
If we diminish its value the maximum of infected people, $i$,  will be augmented.
% Fig 01

We see that the fitting is good and this gives us confidence in the model so far with
the available data up to that date but we should be aware that the system is 
open and so we will be careful with the conclusions. However, as it is shown in 
\cite{Mar2010} a maximum value of infected people was predicted that is 
in agreement with the first observed maximum value.
In the next section we improve the model to 
study the spatial behavior in addition to the temporal behavior.

%%%%%%%%%%%%%%%%% SECTION %%%%%%%%%%%%%%%%%%
\section{Dynamics of populations: the spatial behavior}\label{Spatial_sec}

Let us assume that the infection can be spread over a city or a spatial region that 
occupies a region 
in the $xy$-plane. We write
\begin{eqnarray}
S&=&S(x,y;t)  \nonumber \\
I&=&I(x,y;t) \\
R&=&R(x,y;t) \nonumber
\end{eqnarray}
to indicate that now the populations have a spatial and temporal behavior and we say
that $S(x,y;t)$ is the density of susceptible population in persons per square kilometer,
for example, in a small square centered at $(x,y)$. 
The same interpretation is for $I(x,y;t)$ and $R(x,y;t)$.

The basic idea to obtain the equations of the new SIR model 
that govern the spatial and temporal behavior of the populations
is to use also all the assumptions that lead us to Eqs. (\ref{SIR_eq1}--\ref{SIR_eq3})
to build up the temporal behavior. We model the spatial behavior by adding a
diffusion term in such a way that we take into account the fact that populations
are spread over the region and it takes time for an infection to spread across that
region (see Ref.\ \cite{Murray1989} for other possible diffusion models). 
The evolution equations then are
\begin{eqnarray}
\frac{\partial s}{\partial t} &=& - i s -  (\nabla^2 i) s \label{SIR_spatial_eq1} \\
\frac{\partial i}{\partial t} &=& i s +  (\nabla^2 i) s  - \rho i \label{SIR_spatial_eq2}
\end{eqnarray}
We may notice also that the infected people have the greatest influence on susceptible
persons
that are spatially closest to them. 
We are assuming that infected individuals recover at the same rate at all places.
Spatial coordinates are in units of $(d/Na)^{1/2}$ where $d$ is the diffusion constant
and the time is again in units of $1/Na$, the infection period.
The diffusion effects of the released people do not have any effects on the spatial dynamics
of susceptible population or infected population and its equation is given by 
Eq. (\ref{SIR_eq3}) but with
the new meaning of the populations. We note also that again parameter $\rho$
is the most important parameter of the model.

Let us assume that the city is a square region with sides of size $l$ and we 
solve Eqs. (\ref{SIR_spatial_eq1}) and (\ref{SIR_spatial_eq2}) 
with the boundary conditions: 
\begin{eqnarray}
\hspace{-1.2em}
\frac{\partial s(0,y;t)}{\partial x} = \frac{\partial s(l,y;t)}{\partial x} 
= \frac{\partial i(0,y;t)}{\partial x} = \frac{\partial i(l,y;t)}{\partial x} = 0 , 
\label{SIR_spatial_boundary_eq1}\\
\hspace{-1.2em}
\frac{\partial s(x,0;t)}{\partial y} = \frac{\partial s(x,l;t)}{\partial y} 
= \frac{\partial i(x,0;t)}{\partial y} = \frac{\partial i(x,l;t)}{\partial y} = 0 . 
\label{SIR_spatial_boundary_eq2}
\end{eqnarray}
This means that there is no migration through the boundaries consistent with assumption (e)
of Sec. \ref{SIR_model_sec}.

% Fig 02
\begin{figure}[htbp]
\includegraphics[width=2.07in]{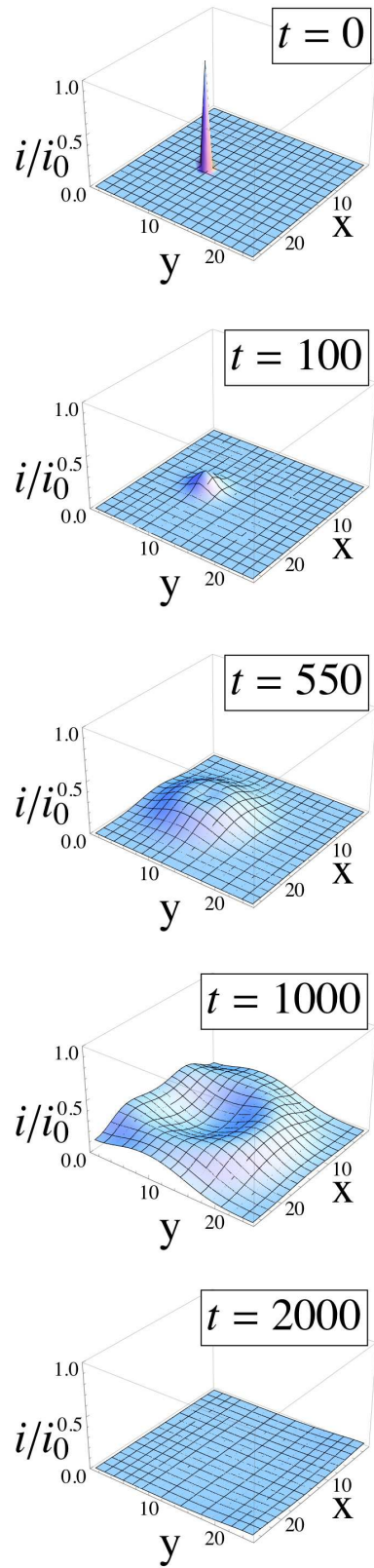} 
\caption{Snapshots of the evolution of the infected density over a region of size $l=25$ in
units of $(d/Na)^{1/2}$. 
We have used $\rho=0.9928$
and $i_0=10^{-4} s_0$.}
\label{fig02_snapshots}
\end{figure}
In Fig. \ref{fig02_snapshots} we show snapshots of the spatial and temporal evolution of the
infected populations for several times. Time is in units of $1/Na$ and the spatial length
is in units of $(d/Na)^{1/2}$. The infected population are given in units of $i_0$,
which is the density of the initial infected people.
Close to the center we introduce an infected population $i_0=10^{-4} s_0$, 
one person in ten thousand is ill in a small area,
and by solving numerically Eqs. (\ref{SIR_spatial_eq1}) and  (\ref{SIR_spatial_eq2})
we evolve the infected and susceptible populations (see 
Refs. \cite{Mar2010b} and \cite{Becerril2008}
for details about how to solve numerically a partial differential equation). 
Parameter $\rho=0.9928$
which is the value we obtained by fitting the Mexican data shown in Fig. \ref{fig01_May_8_plot}.
Also we recall that the other parameters values were for the temporal behavior $Na=62$ and
$s_0=0.99999988$. Then, if we consider that the city is Mexico city modeled with a square
region of sides $l=55$ km in length, we can obtain that the diffusion constant has the value 
$d=300.08$ km$^2/$day. In addition, the population of Mexico city is around 8 million people,
which gives us a mean density of 12,800 people per square unit length. That is way we have
used $i_0=10^{-4} s_0$ for the initial infected density. The spatial and temporal behavior of
the infected population depend on this value or on the $s_0$ parameter. We have selected
$i_0$ density in a consistent way with the value of $s_0$ we obtained by fitting the 
temporal data of the Mexican case.

The infected peak spreads initially, then in a later time diminishes and it forms
an infected wave that propagates to the boundary to finally disappears. It is important
to recall that the diffusion constant is contained in the scaling length $(d/Na)^{1/2}$
so the spatial behavior should be generic. If $\rho$ were greater than $s_0$ no
epidemic develop and the peak diminishes until it vanishes. But if $\rho < s_0$ the
infected density increases and an epidemic develops.

%%%%%%%%%%%%%%% CONCLUSIONS %%%%%%%%%%%%%%%%
\section{Conclusions}\label{conclusions}
We have applied one of the simplest model to describe epidemics, the SIR model.
In this work we use it to model the pandemic by the virus A/H1N1-2009 in the
case of Mexico. One of the main assumptions of the model is to consider that
the population is closed or that it interacts very weakly with its surroundings.
Also, it does not consider that the disease can spread spatially over a region and
that it takes time to spread across that region.
We have improved the SIR model in order to take into account spatial behavior
of the infected populations by adding a diffusion term to the equations and we 
obtain Eqs. (\ref{SIR_spatial_eq1}) and (\ref{SIR_spatial_eq2}).
The diffusion effects of the released people do not have any effects on the dynamics
and we do not consider its equation given by (\ref{SIR_eq3}).
We have used this spatial and temporal SIR
model to study the spatial dynamics of the epidemic in Mexico and assume that
the population is closed which is a very stringent condition but however can give
us insight in the spatial and temporal behavior of an epidemic. That
the system is closed is given by the boundary conditions, Eqs. (\ref{SIR_spatial_boundary_eq1})
and (\ref{SIR_spatial_boundary_eq2}).
The parameters
we use are consistent with the ones obtained by fitting the Mexican data. The
diffusion coefficient is absorbed in the scaling of the spatial length. 
Diffusion can be diminished by public health measures like the ones
the Mexican government took in april 2009.
Finally, the main purpose of this work was to make a first study of an epidemic outbreak
when we have the first data and make predictions about the spatial and temporal
behavior of the epidemic. We think that at this stage we can apply the SIR model
or its variants in spite of their limitations. 
The model have to be modified in order to study the oscillatory
long term behavior of epidemic. A possible model to study this oscillatory
stage was presented already in Ref. \cite{Mar2010}.

One of the main disadvantages of the spatial SIR model we presented in this work is that there
is no migration. To improve our model we need to consider that people travel and therefore
are moving between cities. Possible models are considered in Ref. \cite{Liu} and references
therein. A possibility is a mobility model for residents of $n$ cities who may travel between them
to study the spatial spread of infectious\cite{Arino}. In contrast to our model this is a kind of
spatially discrete model.
\vspace{-1\baselineskip}
\begin{acknowledgments}
This work received financial support by 
CONACYT grant number \ CB-2007-01-84133.
\end{acknowledgments}

%%%%%%%%%%%%%%%%%% BIBLIO %%%%%%%%%%%%%%%%%%

\end{document}